# Koopman-Operator-Based Model Predictive Control for Drag-free Satellite

**Yankai Wang[1] and Ti Chen[1*]**

[1] State Key Laboratory of Mechanics and Control for Aerospace Structures, Nanjing University of Aeronautics and Astronautics, Nanjing, Jiangsu, China

[*]E-mail: chenti@nuaa.edu.cn

**Abstract.** This paper presents a data-driven modelling method for nonlinear dynamics of drag-free satellite based on Koopman operator theory, and a model predictive controller is designed based on the identified model. The nonlinear dynamics of drag-free satellite are identified and controlled based on Sparse Identification of Nonlinear Dynamics (SINDy). Using the manually constructed nonlinear function dictionary as observables, the system approximation is obtained by SINDy algorithm, and a linear Model Predictive Control (MPC) controller is designed for test mass capture based on the SINDy model. Finally, the effectiveness of MPC control is verified by numerical examples.

## 1. Introduction

The constellation of three drag-free satellites provides a viable approach for space-based gravitational wave (GW) detection. The drag-free satellite is a complex multi-body system with two test masses (TMs) housed inside the satellite [1][2]. Space-borne GW missions rely on laser interferometers to measure extremely small relative distance changes between pairs of free-falling TMs. GW signals are extremely weak, and even small non-gravitational disturbances can obscure or distort them. A high-precision drag-free attitude control system is required to ensure that the TMs remain in free-fall and that the satellite accurately follows TMs' motion [3]. Precise control of the drag-free satellite is therefore essential to the success of these space-borne GW detection.

Nowadays, the existing studies on drag-free satellite control often rely on linearizing and decoupling the dynamic model, and design linear controllers for each loop. Vidano et al. [4] proposed an $H_\infty$ controller for drag-free satellite based on a constrained decoupling approach of linearized dynamics. Ma and Wang [5] developed a linear quadratic regulator (LQR) controller with frequency domain constraints. In Reference [6], two Model Predictive Control (MPC) controllers are proposed for TM capture with the command saturations of Wide Range and High Resolution mode. These control methods are typically built upon simplified linearized models, whose accuracy and effectiveness depend strongly on the chosen linearization techniques.

Common linearization approaches include Taylor expansion, piecewise linearization, and orthogonal function approximation. However, these methods have notable limitations. Taylor-based models are only valid near equilibrium points. Piecewise linearization requires extensive model information. Orthogonal function methods involve high computational cost and complicate the controller design. When dealing with highly nonlinear and coupled drag-free satellite dynamics, such linearized approaches may result in performance degradation or fail to capture essential system behaviors. This has motivated researchers to explore alternative modelling and control frameworks that can directly address system nonlinearity without relying on local approximations.

In 1931, Bernard O. Koopman showed that nonlinear dynamical systems can be represented by infinite-dimensional linear operators acting on observable functions in a Hilbert space. The spectral decomposition of the Koopman operator can fully describe the behavior of nonlinear systems. In recent years, Koopman operator theory has attracted renewed attention due to its success in nonlinear system prediction. One key challenge is to obtain a finite-dimensional approximation of the Koopman operator. In some cases, it is possible to derive the operator analytically. Chen et al. [7] used Koopman operator theory to represent attitude dynamics on *SO(*3) as a finite-dimensional linear system by introducing a set of analytical observable functions.

However, in the presence of unknown disturbances or complex system dynamics, it becomes difficult to select appropriate observable functions analytically. In such cases, data-driven methods can be used to discover the underlying dynamic model.

Several data-driven approaches have been developed to approximate the Koopman operator, including dynamic mode decomposition (DMD) [8] and sparse identification of nonlinear dynamics (SINDy). Kaiser et al. [8] showed that DMD-based MPC performs well even for strongly nonlinear systems, when compared with MPC using a fully nonlinear model. The extended DMD (EDMD) algorithm [10][11] improves DMD by using an enhanced observable vector that includes nonlinear functions of the state, improving the accuracy of Koopman approximation for nonlinear systems. SINDy, DMD, and their variants have been successfully applied to systems such as wheeled mobile robots [12], soft robots [13], and wrist rehabilitation robots [14]. In addition, deep learning methods have been introduced to discover suitable observable functions and approximate Koopman operators. Takeishi et al. [15] combined DMD with an autoencoder to learn Koopman operators directly from data. Lusch et al. [16] further improved this framework by introducing an auxiliary network to capture the continuous spectrum.

Koopman operator theory offers a promising framework for modelling and controlling complex systems. In this paper, the finite dimensional approximation of the Koopman operator for drag-free satellite dynamics is obtained using the SINDy algorithm. A SINDy-based MPC is proposed for TM capture. The control performance of the resulting SINDy-MPC controller is evaluated through simulation.

The remainder of this paper is organized as follows. Section 2 introduces the Koopman operator theory, nonlinear dynamics of drag-free satellite and SINDy algorithm. The linear SINDy-MPC algorithm is given in Section 3. In Section 4, the simulation results are provided. Conclusions are drawn in Section 5.

## 2. Koopman Model

### *2.1 Koopman Operator Theory*

The Koopman operator theory provides a method to transform a nonlinear system into an infinite-dimensional linear system. Consider a nonlinear system

$$\dot{\boldsymbol{x}} = \boldsymbol{f}\left(\boldsymbol{x}\right) \tag{1}$$

where $\boldsymbol{x}\left(k\right) \in \mathbb{R}^n$ is the state of the system, and $\boldsymbol{f}$ is the function of the state evolution in the state space.

The Koopman operator $\boldsymbol{K}_t$ is an infinite linear operator. The evolution on the observables $\boldsymbol{g}$ is represented by

$$\boldsymbol{K}_t \boldsymbol{g}(\boldsymbol{x}) = \boldsymbol{g}(\boldsymbol{f}(\boldsymbol{x})) \tag{2}$$

Then the discrete form of nonlinear system with $\Delta t$, one has:

$$\boldsymbol{K}_{\Delta t} \boldsymbol{g}(\boldsymbol{x}(k)) = \boldsymbol{g}(\boldsymbol{f}_{\Delta t}(\boldsymbol{x}(k))) = \boldsymbol{g}(\boldsymbol{x}(k+1)) \tag{3}$$

The linearity of Koopman theory is very appealing, but it is necessary to truncate the infinite-dimensional operator to finite matrix approximation in practice.

*2.2 Nonlinear Dynamics Model of Drag-free Satellite*

This subsection aims to present a concise overview of the dynamics equations of a drag-free satellite. The following reference frames are defined.

Inertial Reference Frame (IRF) $O_I \boldsymbol{X}_I \boldsymbol{Y}_I \boldsymbol{Z}_I$: The origin is centred at Earth's center of mass (CoM). The $\boldsymbol{X}_I$ pointing to the mean Vernal-equinox at epoch J2000, the $\boldsymbol{Z}_I$ is normal to the mean equator at epoch J2000 and $\boldsymbol{Y}_I$ completes the right-handed system.

Satellite Reference Frame (SRF) $O_S \boldsymbol{X}_S \boldsymbol{Y}_S \boldsymbol{Z}_S$: The origin is centred in the satellite's CoM, the $\boldsymbol{X}_S$ is oriented along the angular bisector of the nominal angle between optical assemblies, $\boldsymbol{Z}_S$ is perpendicular to the solar panel and $\boldsymbol{Y}_S$ completes the right-handed system.

Optical Reference Frame (ORFi) $O_{Oi} \boldsymbol{X}_{Oi} \boldsymbol{Y}_{Oi} \boldsymbol{Z}_{Oi}$ (i=1, 2): The origin is centred in the electrostatic suspension cage center, the $\boldsymbol{Z}_{Oi}$ is parallel to the $\boldsymbol{Z}_S$, the $\boldsymbol{X}_{Oi}$ is directed toward the laser link and $\boldsymbol{Y}_{Oi}$ completes the right-handed system.

Test-Mass Reference Frame (MRFi) $O_{Mi} \boldsymbol{X}_{Mi} \boldsymbol{Y}_{Mi} \boldsymbol{Z}_{Mi}$ (i=1, 2): The origin is centred in the TM's CoM, the axis is perpendicular to the surface of the test-mass.

The attitude dynamics for drag-free satellite in the IRF is given as follows:

$$\dot{\boldsymbol{\omega}}_{SI} = -\boldsymbol{J}_S^{-1} \boldsymbol{\omega}_{SI}^{\times} \boldsymbol{J}_S \boldsymbol{\omega}_{SI} + \boldsymbol{J}_S^{-1} \left( \boldsymbol{M}_T + \boldsymbol{D}_S - \sum_{i=1,2} \left( \boldsymbol{R}_{Oi}^S I_{zzi} \ddot{\gamma}_i + \boldsymbol{R}_{Oi}^S \boldsymbol{M}_{Ei} + \boldsymbol{b}_i^{\times} \boldsymbol{R}_{Oi}^S \boldsymbol{F}_{Ei} \right) \right) \tag{4}$$

where $\boldsymbol{\omega}_{SI}$ is the angular velocity of the satellite in the IRF, $\boldsymbol{J}_S$ denotes the inertia matrix of the satellite, $\boldsymbol{M}_T$ is the input torque of the thrusters, $\boldsymbol{D}_S$ is the total external disturbance, $\boldsymbol{R}_{Oi}^S$ is the rotation matrix from ORFi to SRF, $I_{zzi}$ is inertia of i-th Moving Optical Sub-Assembly (MOSA), $\gamma_i$ is the angle between SRF and i-th MOSA, $\boldsymbol{M}_{Ei}$ and $\boldsymbol{F}_{Ei}$ are the electrostatic torque and force, respectively. $\boldsymbol{b}_i$ is the position of the cage center in the SRF, $\boldsymbol{a}^{\times} : \mathbb{R}^3 \to \mathfrak{so}(3)$ is defined as the skew-symmetric matrix.

The dynamics for TM in the ORF are given as follows:

$$\begin{aligned} \ddot{\boldsymbol{r}}_{MOi} = {} & \boldsymbol{R}_I^O \boldsymbol{G}_{\Delta i} + m_{Mi}^{-1} (\boldsymbol{F}_{Ei} + \boldsymbol{d}_{Mi} + \boldsymbol{F}_{Sti}) - m_{Mi}^{-1} \boldsymbol{R}_S^{Oi} (\boldsymbol{F}_T + \boldsymbol{d}_S) + m_S^{-1} \sum_{j=1,2} \boldsymbol{R}_{Oj}^{Oi} \boldsymbol{F}_{Ej} \\ & - \boldsymbol{R}_S^{Oi} \Omega(\boldsymbol{\omega}_{SI}) \boldsymbol{b}_{Si} - \Omega(\boldsymbol{\omega}_{OiI}) \boldsymbol{b}_{Mi} - \Omega(\boldsymbol{\omega}_{OiI}) \boldsymbol{r}_{MOi} - 2 \boldsymbol{\omega}_{OiI}^{\times} \dot{\boldsymbol{r}}_{MOi} \end{aligned} \tag{5}$$

$$\begin{aligned} \dot{\boldsymbol{\omega}}_{MOi} = {} & -\boldsymbol{J}_{Mi}^{-1} \boldsymbol{\omega}_{MiI}^{\times} \boldsymbol{J}_{Mi} \boldsymbol{\omega}_{MiI} + \boldsymbol{J}_{Mi}^{-1} \boldsymbol{R}_{Oi}^{Mi} (\boldsymbol{M}_{Ei} + \boldsymbol{D}_{Mi} + \boldsymbol{M}_{Sti}) \\ & - \boldsymbol{R}_{Oi}^{Mi} \boldsymbol{\omega}_{\gamma i} - \boldsymbol{R}_{Oi}^{Mi} \dot{\boldsymbol{\omega}}_{\gamma i} - \dot{\boldsymbol{R}}_S^{Mi} \boldsymbol{\omega}_{SI} - \boldsymbol{R}_S^{Mi} \dot{\boldsymbol{\omega}}_{SI} \end{aligned} \tag{6}$$

where $\boldsymbol{r}_{MOi}$ is the relative position of i-th TM, $\boldsymbol{G}_{\Delta i}$ is the gravity gradient, $m_{Mi}$ is the TM's mass, $\boldsymbol{F}_{Sti}$ and $\boldsymbol{M}_{Sti}$ are the stiffness force and torque, respectively, $\boldsymbol{F}_T$ is the control vector pf the satellite, $\boldsymbol{d}_S$ is the disturbance force, $\Omega(\boldsymbol{x})$ is defined as $\dot{\boldsymbol{x}}^{\times} + \boldsymbol{x}^{\times} \boldsymbol{x}^{\times}$ for $\boldsymbol{x} \in \mathbb{R}^3$. $\boldsymbol{b}_{Si}$ is the position of i-th MOSA pivot in the SRF, $\boldsymbol{b}_{Mi}$ is the vector from i-th pivot to i-th cage center, $\boldsymbol{\omega}_{OiI}$ is the angular velocity of i-th MOSA in the IRF. $\boldsymbol{\omega}_{MOi}$ is the angular velocity of i-th TM in the ORF, $\boldsymbol{J}_{Mi}$ is the inertia matrix of $i$-th TM, $\boldsymbol{\omega}_{MiI}$ is the angular velocity of $i$-th TM in the IRF, $\boldsymbol{d}_{Mi}$ and $\boldsymbol{D}_{Mi}$ are the disturbance acting on i-th TM, $\boldsymbol{\omega}_{\gamma i}$ is the angular velocity of i-th MOSA.

The rotation dynamics for MOSA is given as follows:

$$\ddot{\zeta}_i = -2\omega_{Ni}\xi_i\dot{\zeta}_i - \omega_{Ni}^2\zeta_i + \frac{M_{MOSAi}}{I_{zzi}} - \frac{\boldsymbol{e}_3^{\mathrm{T}}\boldsymbol{M}_{Ei}}{I_{zzi}} - \boldsymbol{e}_3^T\boldsymbol{R}_S^{Oi}\dot{\boldsymbol{\omega}}_{SI} + \frac{D_{\zeta i}}{I_{zzi}} \tag{7}$$

where $\zeta_i$ is the angle of $i$-th MOSA w.r.t the nomial positon, $\omega_{Ni}$ is the natural angular frequency, $\xi_i$ is the damping ratio, $\boldsymbol{M}_{MOSAi}$ is the total toque, $D_{\zeta i}$ is the disturbance, $\boldsymbol{e}_3 = \begin{bmatrix}0 & 0 & 1\end{bmatrix}^{\mathrm{T}}$.

*2.3 Data-Driven Finite-dimensional Approximation of the Koopman*

SINDy is an effective method to construct a finite-dimensional approximation of the Koopman operator. Consider a nonlinear controlled system:

$$\frac{\mathrm{d}}{\mathrm{d}t}\boldsymbol{x} = \boldsymbol{f}(\boldsymbol{x}(k),\boldsymbol{u}(k)) \tag{8}$$

where $\boldsymbol{u}(k) \in \mathbb{R}^m$ is the control input.

The time-series data is collected at several times $k$, $k+1$, $\cdots$, $k+p$ and formed into a data matrix:

$$\boldsymbol{X} = \begin{bmatrix}\boldsymbol{x}(k) & \boldsymbol{x}(k+1) & \cdots & \boldsymbol{x}(k+p)\end{bmatrix}^{\mathrm{T}} \tag{9}$$

Similar matrices are formed as follows:

$$\dot{\boldsymbol{X}} = \begin{bmatrix}\dot{\boldsymbol{x}}(k) & \dot{\boldsymbol{x}}(k+1) & \cdots & \dot{\boldsymbol{x}}(k+p)\end{bmatrix}^{\mathrm{T}} \tag{10}$$

$$\boldsymbol{U} = \begin{bmatrix}\boldsymbol{u}(k) & \boldsymbol{u}(k+1) & \cdots & \boldsymbol{u}(k+p)\end{bmatrix}^{\mathrm{T}} \tag{11}$$

Construct an augmented library:

$$\boldsymbol{\Theta}(\boldsymbol{X},\boldsymbol{U}) = \begin{bmatrix}\theta_1(\boldsymbol{X},\boldsymbol{U}) & \theta_2(\boldsymbol{X},\boldsymbol{U}) & \cdots\end{bmatrix} \tag{12}$$

where $\theta_i$ is candidate nonlinear functions of $\boldsymbol{X}$ and $\boldsymbol{U}$.

The dynamical system (8) can be represented as follows:

$$\boldsymbol{f}(\boldsymbol{X},\boldsymbol{U}) = \dot{\boldsymbol{X}} = \boldsymbol{\Theta}(\boldsymbol{X},\boldsymbol{U})\boldsymbol{\Xi} \tag{13}$$

where $\boldsymbol{\Xi} = \begin{bmatrix}\xi_1 & \xi_2 & \cdots & \xi_k\end{bmatrix}$ is the sparse matrix of coefficients, and each column $\xi_k$ is a vector of coefficients determining the active terms.

The parsimonious model can be identified using a convex $\ell_1$-regularized saprse regression:

$$\xi_k = \arg\min_{\xi_k'} \left\|\dot{\boldsymbol{X}}_k - \boldsymbol{\Theta}(\boldsymbol{X},\boldsymbol{U})\xi_k'\right\| + \lambda\left\|\xi_k'\right\|_1 \tag{14}$$

where $\dot{\boldsymbol{X}}_k$ is the k-th column of $\dot{\boldsymbol{X}}$, and $\lambda$ is a sparsity-promoting knob.

The nonlinear observables $\boldsymbol{\Psi} \in \mathbb{R}^N$, control coupling function $\boldsymbol{\Psi}_u \in \mathbb{R}^M$, and higher-order compensation observables $\bar{\boldsymbol{\Psi}} \in \mathbb{R}^{\bar{N}}$ that constitute the Koopman lift space, are selected as follows:

$$\boldsymbol{\Psi}(k) = \begin{bmatrix}\psi_1(\boldsymbol{x}(k)) & \psi_2(\boldsymbol{x}(k)) & \cdots & \psi_N(\boldsymbol{x}(k))\end{bmatrix} \tag{15}$$

$$\boldsymbol{\Psi}_u(k) = \begin{bmatrix}\psi_{u1}(\boldsymbol{x}(k),\boldsymbol{u}(k)) & \psi_{u2}(\boldsymbol{x}(k),\boldsymbol{u}(k)) & \cdots & \psi_{uM}(\boldsymbol{x}(k),\boldsymbol{u}(k))\end{bmatrix} \tag{16}$$

$$\bar{\boldsymbol{\Psi}}(k) = \begin{bmatrix}\bar{\psi}_1(\boldsymbol{x}(k)) & \bar{\psi}_2(\boldsymbol{x}(k)) & \cdots & \bar{\psi}_{\bar{N}}(\boldsymbol{x}(k))\end{bmatrix} \tag{17}$$

where $\psi_i$, $i \in [1,N]$, $\psi_{ui}$, $i \in [1,M]$ and $\bar{\psi}_i$, $i \in [1,\bar{N}]$ are the candidate observation function.

The augmented library can be designed as

$$\boldsymbol{\Theta} = \begin{bmatrix} \boldsymbol{\Psi} & \bar{\boldsymbol{\Psi}} & \boldsymbol{\Psi}_u \end{bmatrix} \tag{18}$$

A generalized linear model in the lifted space can be obtained using a sparse regression algorithm.

$$\dot{\boldsymbol{X}}_{lift} = \boldsymbol{\Theta}\boldsymbol{\Xi} \tag{19}$$

where $\dot{\boldsymbol{X}}_{lift} = \begin{bmatrix} \dot{\boldsymbol{\Psi}}(k) & \dot{\boldsymbol{\Psi}}(k+1) & \cdots & \dot{\boldsymbol{\Psi}}(k+p) \end{bmatrix}$.

The generalized linear model of the nonlinear system is then expressed in state-space form:

$$\dot{\boldsymbol{\Psi}}^{\mathrm{T}} = \boldsymbol{A}\boldsymbol{\Psi}^{\mathrm{T}} + \bar{\boldsymbol{A}}\bar{\boldsymbol{\Psi}}^{\mathrm{T}} + \boldsymbol{B}\boldsymbol{\Psi}_u^{\mathrm{T}} \tag{20}$$

where $\boldsymbol{A}\boldsymbol{\Psi}^{\mathrm{T}} + \boldsymbol{B}\boldsymbol{\Psi}_u^{\mathrm{T}}$ represents the linear component of the model, while $\bar{\boldsymbol{A}}\bar{\boldsymbol{\Psi}}^{\mathrm{T}}$ is the higher-order terms. By neglecting the higher-order terms in $\bar{\boldsymbol{A}}\bar{\boldsymbol{\Psi}}^{\mathrm{T}}$, Equation (20) simplifies to a linear system:

$$\dot{\boldsymbol{\Psi}}^{\mathrm{T}} = \boldsymbol{A}\boldsymbol{\Psi}^{\mathrm{T}} + \boldsymbol{B}\boldsymbol{\Psi}_u^{\mathrm{T}} \tag{21}$$

This results in a finite-dimensional approximation of the Koopman operator for the nonlinear dynamics. The system (21) is discretized to obtain the discrete state-space equations in the lifted space:

$$\boldsymbol{\chi}_{k+1} = \boldsymbol{A}_D\boldsymbol{\chi}_k + \boldsymbol{B}_D\bar{\boldsymbol{u}}_k \tag{22}$$

where $\chi$ represents the state of the model in the lifted space formed by the vector $\boldsymbol{\Psi}$, $\bar{\boldsymbol{u}}$ is the control input of the discrete system.

Sparse regression algorithms include Least Absolute Shrinkage and Selection Operator (LASSO) and Sequential Thresholded Least-Squares (STLS). In this study, the STLS algorithm is adopted to compute the SINDy model. The fundamental idea of STLS is to enforce sparsity in the least-squares solution by thresholding small coefficients based on a sparsity-promoting threshold. The detailed computational procedure is as follows:

1. Obtain data snapshots $\boldsymbol{X}$ and $\dot{\boldsymbol{X}}$ through simulations or experiments and compute the corresponding lifted data snapshots $\boldsymbol{\Theta}(\boldsymbol{X},\boldsymbol{U})$ and $\dot{\boldsymbol{X}}_{lift}$.

2. Initialize the optimal fitting operator $\boldsymbol{\Xi} = \boldsymbol{\Theta} \backslash \dot{\boldsymbol{X}}_{lift}$ using the least-squares method.

3. Set all coefficients in $\boldsymbol{\Xi}$ smaller than the sparsity threshold $\lambda$ to zero. Construct the matrix $\boldsymbol{B}_i$, where nonzero elements in $\boldsymbol{\Xi}$ are marked as 1, and all other elements are set to 0.

4. Recompute the sparse optimal fitting operator $\boldsymbol{\Xi}$ using the least-squares method, considering only the nonzero elements' corresponding data. Specifically, $\xi_k = \left(\boldsymbol{\Theta}(\boldsymbol{X},\boldsymbol{U}) \cdot \boldsymbol{B}_{ik}\right) \backslash \dot{\boldsymbol{X}}_{lift}$ is used, where $\xi_k$ represents the $k$-th column of $\boldsymbol{\Xi}$, $\boldsymbol{B}_{ik}$ represents the $k$-th column of $\boldsymbol{B}_i$, and $\cdot$ denotes the element-wise multiplication operation.

5. Repeat steps 2–4 until the maximum number of iterations is reached or the solution converges.

## 3. Model Predictive Control

Model predictive control (MPC) computes the control input at each time step by solving an optimization problem. This problem is defined over a limited prediction horizon, using system information available at the beginning of the horizon [17]. The optimization is subject to constraints on control inputs and state variables. It aims to minimize a cost function, such as energy consumption or tracking error. The main advantage of MPC is its ability to consider both future behavior and system constraints when determining the optimal control sequence.

Linear MPC solves a convex quadratic programming (QP) problem, which can be done efficiently. In contrast, nonlinear MPC involves solving a nonconvex problem, which requires significantly more computation. By combining the Koopman linearized model with MPC, it is possible to avoid nonconvex optimization and achieve efficient control of nonlinear systems.

In each control step, MPC solves the optimization problem is formulated as

$$\underset{\boldsymbol{u}_k}{\text{minimize}} \quad J = \sum_{k=0}^{N_p-1} \boldsymbol{e}_k^T \boldsymbol{Q} \boldsymbol{e}_k + \bar{\boldsymbol{u}}_k^T \boldsymbol{R} \bar{\boldsymbol{u}}_k + \Delta \bar{\boldsymbol{u}}_k^T \boldsymbol{S} \Delta \bar{\boldsymbol{u}}_k \tag{23}$$

$$\text{subject to:} \; \boldsymbol{\chi}_{k+1} = \boldsymbol{A}_D \boldsymbol{\chi}_k + \boldsymbol{B}_D \bar{\boldsymbol{u}}_k, \; k = 0, \cdots, N_p - 1 \tag{24}$$

$$\boldsymbol{u}_{\min} \le \bar{\boldsymbol{u}}_k \le \boldsymbol{u}_{\max} \tag{25}$$

where $N_p$ is the the prediction horizon, $k$ is the prediction step, $\boldsymbol{Q}$ $\boldsymbol{R}$ and $\boldsymbol{S}$ are diagonal weighting matrices. $\boldsymbol{e}_k$ is the state error, $\Delta \bar{\boldsymbol{u}}_k = \bar{\boldsymbol{u}}_k - \bar{\boldsymbol{u}}_{k-1}$ denotes the input increment. $\boldsymbol{u}_{\max}$ and $\boldsymbol{u}_{\min}$ are the upper and lower bounds of control input sequence.

The SINDy-based model predictive control algorithm proceeds as follows:

1. At the current time step $k$, measure the state $\boldsymbol{x}_k$ of the nonlinear system. Use the lifting function $\boldsymbol{\Theta}$ to map $\boldsymbol{x}_k$ into the high-dimensional lifted space, to yield the initial value $\boldsymbol{\chi}_0$ for the prediction model.

2. Solve the above optimization problem to obtain the optimal control sequence $\bar{\boldsymbol{U}}$.

3. Extract the first element $\bar{\boldsymbol{u}}_k$ of the optimal control sequence $\bar{\boldsymbol{U}}$.

4. Compute the actual control torque $\boldsymbol{u}_k = \boldsymbol{T}^{-1}(\boldsymbol{x}) \bar{\boldsymbol{u}}_k$ using the state-dependent transformation matrix $\boldsymbol{T}(\boldsymbol{x})$.

5. Apply $\boldsymbol{u}_k$ to the actual system to obtain the next state $\boldsymbol{x}_{k+1}$, then return to step 1 and repeat the process.

## 4. Numerical Verification

*4.1 Data Collection and Koopman Training*

Based on the nonlinear dynamics of the drag-free satellite described in Equations (4)–(7), numerical simulations are conducted in MATLAB to generate training data. The satellite parameters used in the simulation are listed in Table 1.

The simulation is performed using the ode45 solver, and the system states are sampled at intervals of 0.1 s. The random initial conditions are given as follows:

$$\boldsymbol{\theta}_{SI} \in \begin{bmatrix} \theta_{SIx} & \theta_{SIy} & \theta_{SIz} \end{bmatrix}^{\mathrm{T}}, \; \forall j \in \{x, y, z\}, \; \theta_{SIj} \overset{\text{i.i.d}}{\sim} \mathcal{U}(-10^{-8}, 10^{-8}) \tag{26}$$

$$\boldsymbol{\omega}_{SIi} \in \begin{bmatrix} \omega_{SIix} & \omega_{SIiy} & \omega_{SIiz} \end{bmatrix}^{\mathrm{T}}, \; \forall j \in \{x, y, z\}, \; \omega_{SIij} \overset{\text{i.i.d}}{\sim} \mathcal{U}(-10^{-8}, 10^{-8}) \tag{27}$$

$$\forall i \in \{1,2\}, \; \boldsymbol{r}_{MOi} \in \begin{bmatrix} r_{MOix} & r_{MOiy} & r_{MOiz} \end{bmatrix}^{\mathrm{T}}, \; \forall j \in \{x, y, z\}, \; r_{MOij} \overset{\text{i.i.d}}{\sim} \mathcal{U}(-10^{-7}, 10^{-7}) \tag{28}$$

$$\forall i \in \{1,2\}, \; \dot{\boldsymbol{r}}_{MOi} \in \begin{bmatrix} \dot{r}_{MOix} & \dot{r}_{MOiy} & \dot{r}_{MOiz} \end{bmatrix}^{\mathrm{T}}, \; \forall j \in \{x, y, z\}, \; \dot{r}_{MOij} \overset{\text{i.i.d}}{\sim} \mathcal{U}(-10^{-8}, 10^{-8}) \tag{29}$$

$$\forall i \in \{1,2\}, \; \boldsymbol{\theta}_{MOi} \in \begin{bmatrix} \theta_{MOix} & \theta_{MOiy} & \theta_{MOiz} \end{bmatrix}^{\mathrm{T}}, \; \forall j \in \{x, y, z\}, \; \theta_{MOij} \overset{\text{i.i.d}}{\sim} \mathcal{U}(-10^{-5}, 10^{-5}) \tag{30}$$

$$\forall i \in \{1,2\}, \; \boldsymbol{\omega}_{MOi} \in \begin{bmatrix} \omega_{MOix} & \omega_{MOiy} & \omega_{MOiz} \end{bmatrix}^{\mathrm{T}}, \; \forall j \in \{x, y, z\}, \; \omega_{MOij} \overset{\text{i.i.d}}{\sim} \mathcal{U}(-10^{-7}, 10^{-7}) \tag{31}$$

$$\forall i \in \{1,2\},\ \zeta_i \overset{\text{i.i.d}}{\sim} \mathcal{U}(-10^{-8}, 10^{-8}) \tag{32}$$

$$\forall i \in \{1,2\},\ \dot{\zeta}_i \overset{\text{i.i.d}}{\sim} \mathcal{U}(-10^{-10}, 10^{-10}) \tag{33}$$

The control input is defined as a sinusoidal excitation in the form:

$$\boldsymbol{u}(t) = \boldsymbol{A} \circ \boldsymbol{a} \circ \sin(\boldsymbol{b}t + \boldsymbol{c}) \tag{34}$$

where $\boldsymbol{A} = \begin{bmatrix} \boldsymbol{I}_3 \times 10^{-7} & \boldsymbol{I}_3 \times 2 \times 10^{-5} & \boldsymbol{I}_3 \times 10^{-7} & \boldsymbol{I}_3 \times 3 \times 10^{-9} & \boldsymbol{I}_3 \times 10^{-7} & \boldsymbol{I}_3 \times 3 \times 10^{-9} & \boldsymbol{I}_2 \times 10^{-9} \end{bmatrix}^{\mathrm{T}}$, $a_i \sim \mathcal{U}(-1,1)$, $b_i \sim \mathcal{U}(0,5)$ and $c_i \sim \mathcal{U}(0,2\pi)$ for $i = 1,2,\cdots,20$, and $\boldsymbol{I}_3 = \begin{bmatrix} 1 & 1 & 1 \end{bmatrix}$.

**Table 1.** Parameters of the drag-free satellite.

| Parameter | Value | Unit |
|---|---|---|
| $m_S$ | 1500 | kg |
| $\boldsymbol{J}_S$ | diag{800,800,1000} | kg·m$^2$ |
| $m_{M1}, m_{M2}$ | 1.9369 | kg |
| $\boldsymbol{J}_{M1}, \boldsymbol{J}_{M2}$ | diag{6.9,6.9,6.9}×10$^{-4}$ | kg·m$^2$ |
| $\omega_{N1}, \omega_{N2}$ | 72.76 | rad/s |
| $\xi_1, \xi_2$ | 0.0323 | / |
| $\boldsymbol{b}_{S1}$ | [0.1074 0.3216 0]$^{\mathrm{T}}$ | m |
| $\boldsymbol{b}_{S2}$ | [0.1074 -0.3216 0]$^{\mathrm{T}}$ | m |
| $\boldsymbol{b}_{M1}, \boldsymbol{b}_{M2}$ | [0.25 0 0]$^{\mathrm{T}}$ | m |

A total of 200 trajectories are generated under the above random initial conditions and sinusoidal control inputs, each with a duration of 50 s. The derivative data required by the SINDy algorithm are computed using the fourth-order central difference method. Among all data sets, 150 data sets are used for training, and the remaining 50 data sets are used for validation.

In Reference [5], controllers are designed separately for the TM capture loop and the satellite attitude control loop. Based on this structure, sparse identification is applied to model the two dynamic subsystems independently. The polynomial of the state of satellite attitude dynamics are chosen as the nonlinear observables:

$$\boldsymbol{\Psi}_1(k) = \begin{bmatrix} \boldsymbol{\theta}_{SI} & \zeta_1 & \zeta_2 & \boldsymbol{\omega}_{SI} & \dot{\zeta}_1 & \dot{\zeta}_2 & \boldsymbol{\theta}_{SI} \otimes \boldsymbol{\omega}_{SI} & \phi_2(\boldsymbol{\omega}_{SI}) & \dot{\zeta}_1^2 & \dot{\zeta}_2^2 \end{bmatrix} \tag{35}$$

$$\boldsymbol{\Psi}_{u1} = \begin{bmatrix} \boldsymbol{M}_T & M_{OA1} & M_{OA2} \end{bmatrix} \tag{36}$$

where operator $\otimes$ denote the Kronecker product, and $\phi_2(\boldsymbol{\omega}_{SI})$ is the 2-nd order polynomial expansion. The nonlinear observables of TM dynamics system are chosen as:

$$\boldsymbol{\Psi}_2(k)=\left[\boldsymbol{r}_{MO} \quad \boldsymbol{\theta}_{MO} \quad \boldsymbol{r}_{MO2} \quad \boldsymbol{\theta}_{MO2} \quad \dot{\boldsymbol{r}}_{MO} \quad \boldsymbol{\omega}_{MO} \quad \dot{\boldsymbol{r}}_{MO2} \quad \boldsymbol{\omega}_{MO2} \quad \boldsymbol{r}_{MO}\otimes\dot{\boldsymbol{r}}_{MO} \right.$$
$$\left. \boldsymbol{\theta}_{MO}\otimes\boldsymbol{\omega}_{MO} \quad \boldsymbol{r}_{MO}\otimes\dot{\boldsymbol{r}}_{MO} \quad \boldsymbol{\theta}_{MO}\otimes\boldsymbol{\omega}_{MO} \quad \phi_2(\dot{\boldsymbol{r}}_{MO}) \quad \phi_2(\boldsymbol{\omega}_{MO}) \quad \phi_2(\dot{\boldsymbol{r}}_{MO2}) \quad \phi_2(\boldsymbol{\omega}_{MO2})\right] \tag{37}$$

$$\boldsymbol{\Psi}_{u2}=\left[\boldsymbol{F}_{E1} \quad \boldsymbol{M}_{E1} \quad \boldsymbol{F}_{E2} \quad \boldsymbol{M}_{E2}\right] \tag{38}$$

The sparsity-promoting knob is chosen as $\lambda = 10^{-13}$. To evaluate the prediction accuracy of the Koopman linearization model, simulations are performed under 50 sets of random initial conditions. The errors between the model predictions and the numerical solutions indicate the prediction accuracy. And the prediction error is represented by the average error of each component of each element of the vector. Taking the satellite attitude $\boldsymbol{\theta}_{SI}$ as an example, the

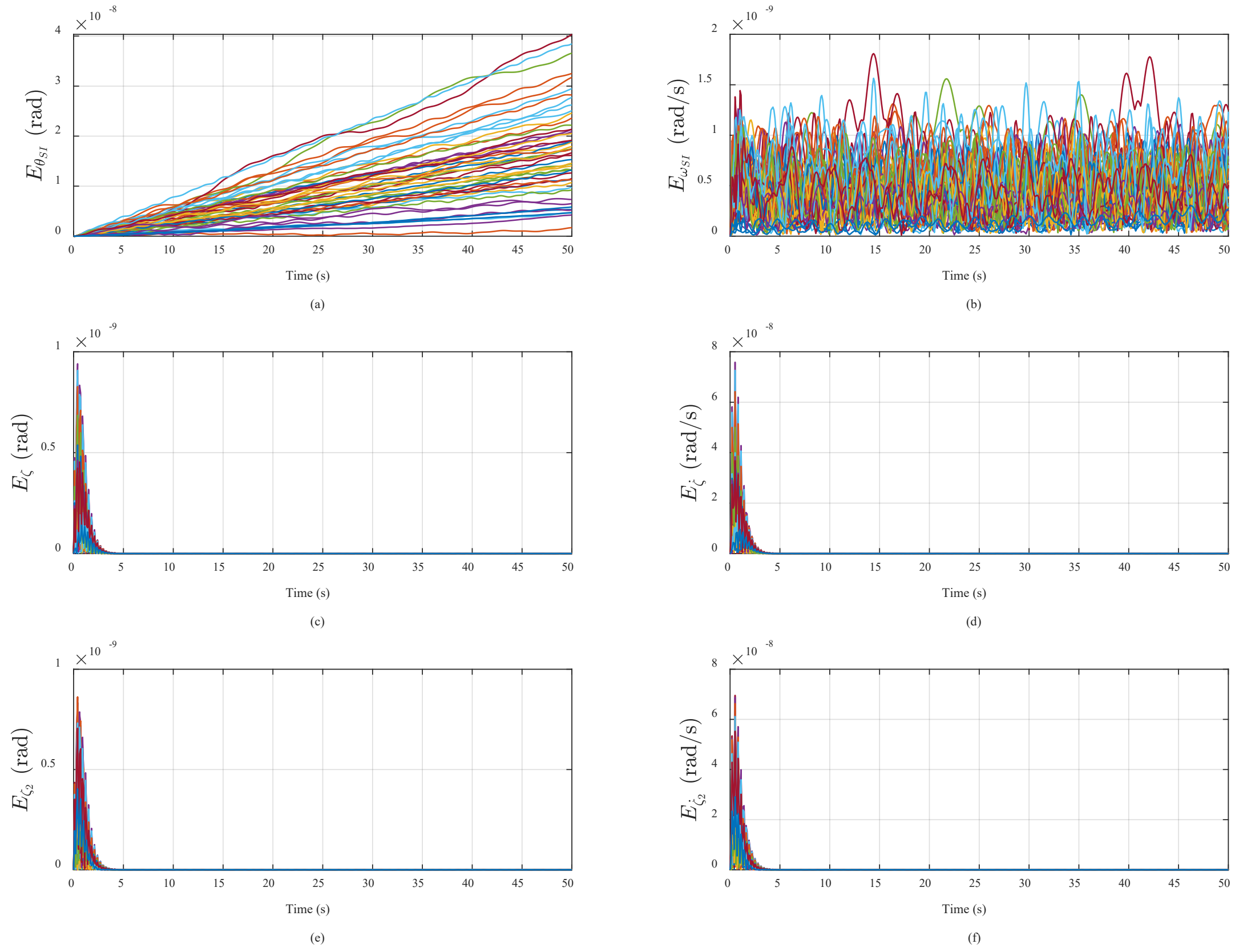

**Figure 1.** Approximation errors of satellite attitude and MOSA's rotation angle with random initial conditions.

prediction error is defined as follows:

$$E_{\theta_{SI}} = \frac{1}{3}\sum_{i=1}^{3}\left|\boldsymbol{\theta}_{SIr}(i) - \boldsymbol{\theta}_{SIp}(i)\right| \tag{39}$$

The prediction errors of the SINDy model for drag-free satellite over a 50 s are shown in Figures 1 and 2. In practice, the control system of a drag-free satellite operates at a frequency of 10 Hz. By appropriately selecting the prediction horizon of the MPC controller, the SINDy model can meet the control accuracy requirements for TM capture control.

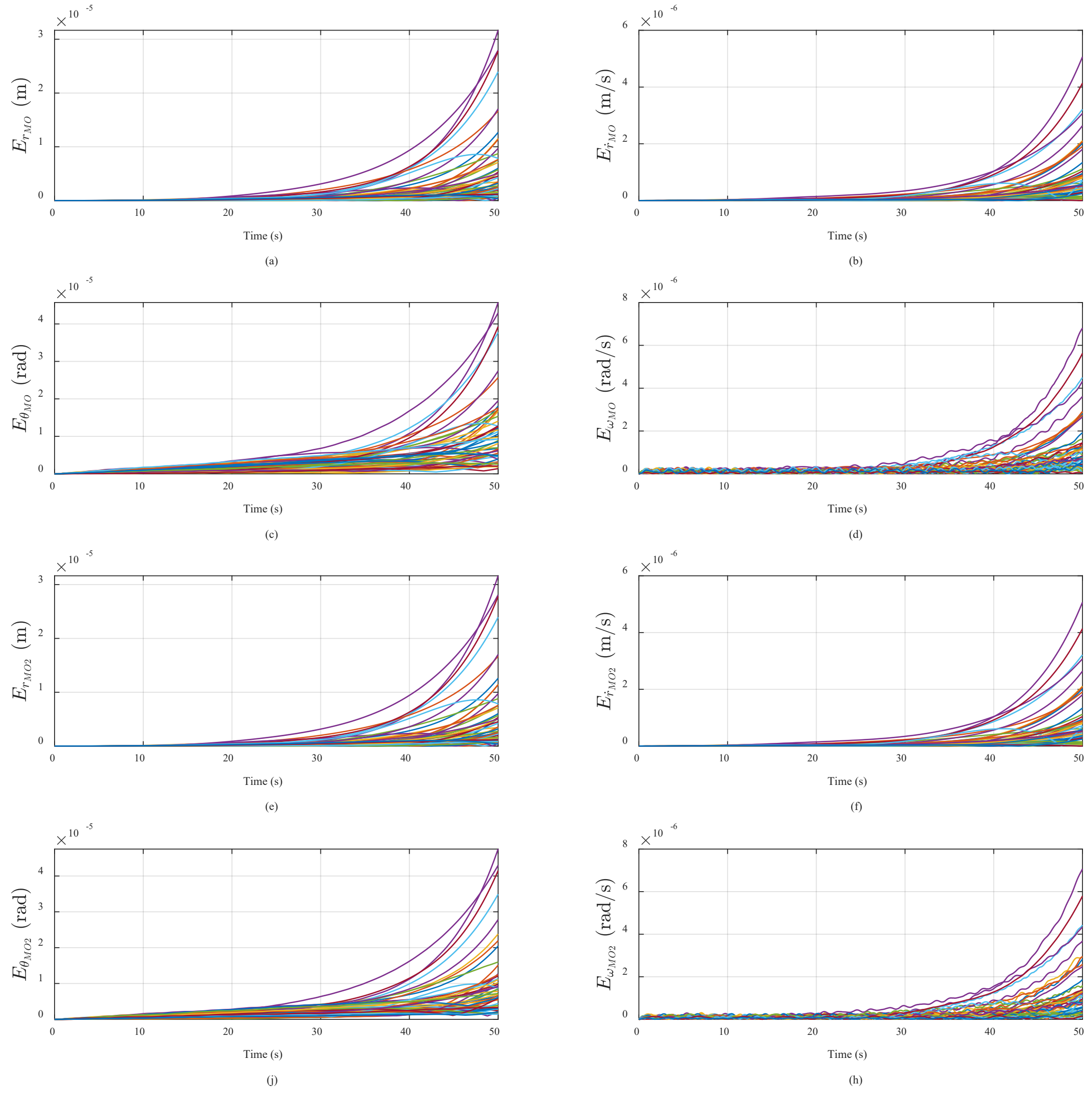

**Figure 2.** Approximation errors of TM' motion with random initial conditions.

*4.2 Controller Performance using Koopman model*

Based on the SINDy model and the SINDy-based MPC controller developed in Section 3, the control accuracy is evaluated through simulation. The initial states of the TM capture mode are given in the Table 2.

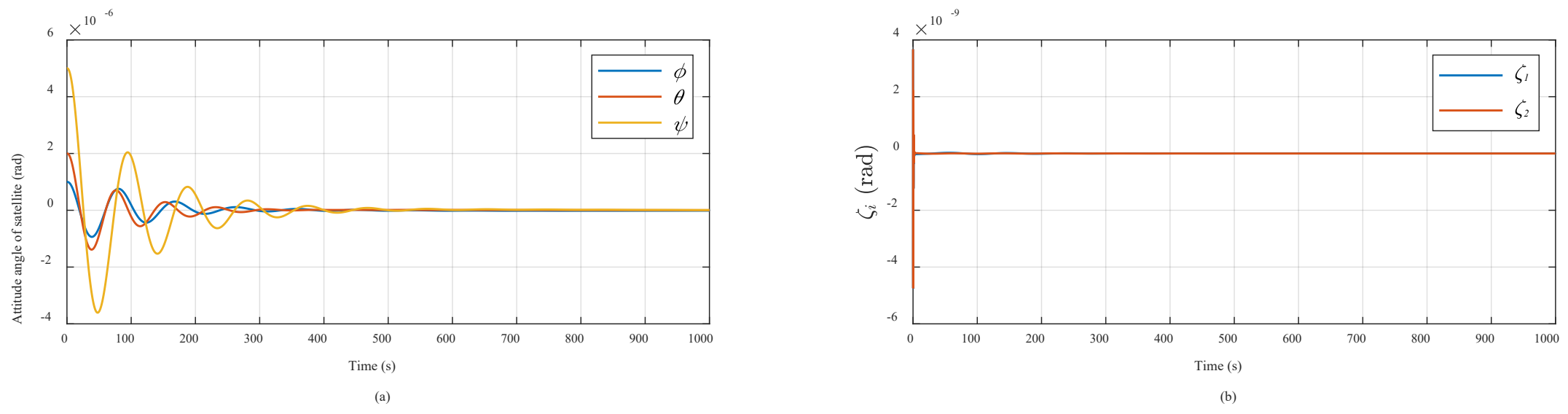

**Figure 3.** The angles of the satellite and MOSAs.

The prediction horizon is chosen as $N_p = 50$, and the control horizon $N_c = 1$. The weighting matrices $\boldsymbol{Q} = \left[ 600 \times \boldsymbol{I}_3 \quad 10 \times \boldsymbol{I}_2 \quad \boldsymbol{0}_{1\times24} \quad 0.005 \times \boldsymbol{I}_3 \quad 0.0004 \times \boldsymbol{I}_3 \quad 0.005 \times \boldsymbol{I}_3 \quad 0.0004 \times \boldsymbol{I}_3 \quad \boldsymbol{0}_{1\times72} \right]$, and $\boldsymbol{R} = \left[ 18 \times \boldsymbol{I}_3 \quad 20 \times \boldsymbol{I}_2 \quad 0.01 \times \boldsymbol{I}_3 \quad 0.001 \times \boldsymbol{I}_3 \quad 0.01 \times \boldsymbol{I}_3 \quad 0.001 \times \boldsymbol{I}_3 \right]$. The upper bounds of control input $\boldsymbol{u}_{\max} = \left[ 2 \times 10^{-5} \quad 1 \times 10^{-6} \right]$, and the lower bounds $\boldsymbol{u}_{\min} = \left[ -2 \times 10^{-5} \quad -1 \times 10^{-6} \right]$.

**Table 2.** Initial states of the drag-free satellite.

| Parameter | Value | Unit |
|---|---|---|
| $\boldsymbol{\theta}_{SI}$ | $[1\ 2\ 5]^{\mathrm{T}}$ | μrad |
| $\boldsymbol{\omega}_{SI}$ | $[0\ 0\ 0]^{\mathrm{T}}$ | μrad/s |
| $\boldsymbol{r}_{MO1}, \boldsymbol{r}_{MO2}$ | $[200\ 200\ 200]^{\mathrm{T}}$ | μm |
| $\dot{\boldsymbol{r}}_{MO1}, \dot{\boldsymbol{r}}_{MO2}$ | $[5\ 5\ 5]^{\mathrm{T}}$ | μm/s |
| $\boldsymbol{\theta}_{MO1}, \boldsymbol{\theta}_{MO2}$ | $[2\ 2\ 2]^{\mathrm{T}}$ | mrad |
| $\boldsymbol{\omega}_{MO1}, \boldsymbol{\omega}_{MO2}$ | $[600\ 600\ 600]^{\mathrm{T}}$ | μrad/s |
| $\zeta_1, \zeta_2$ | 0.01 | μrad |
| $\dot{\zeta}_1, \dot{\zeta}_2$ | 0 | μrad/s |

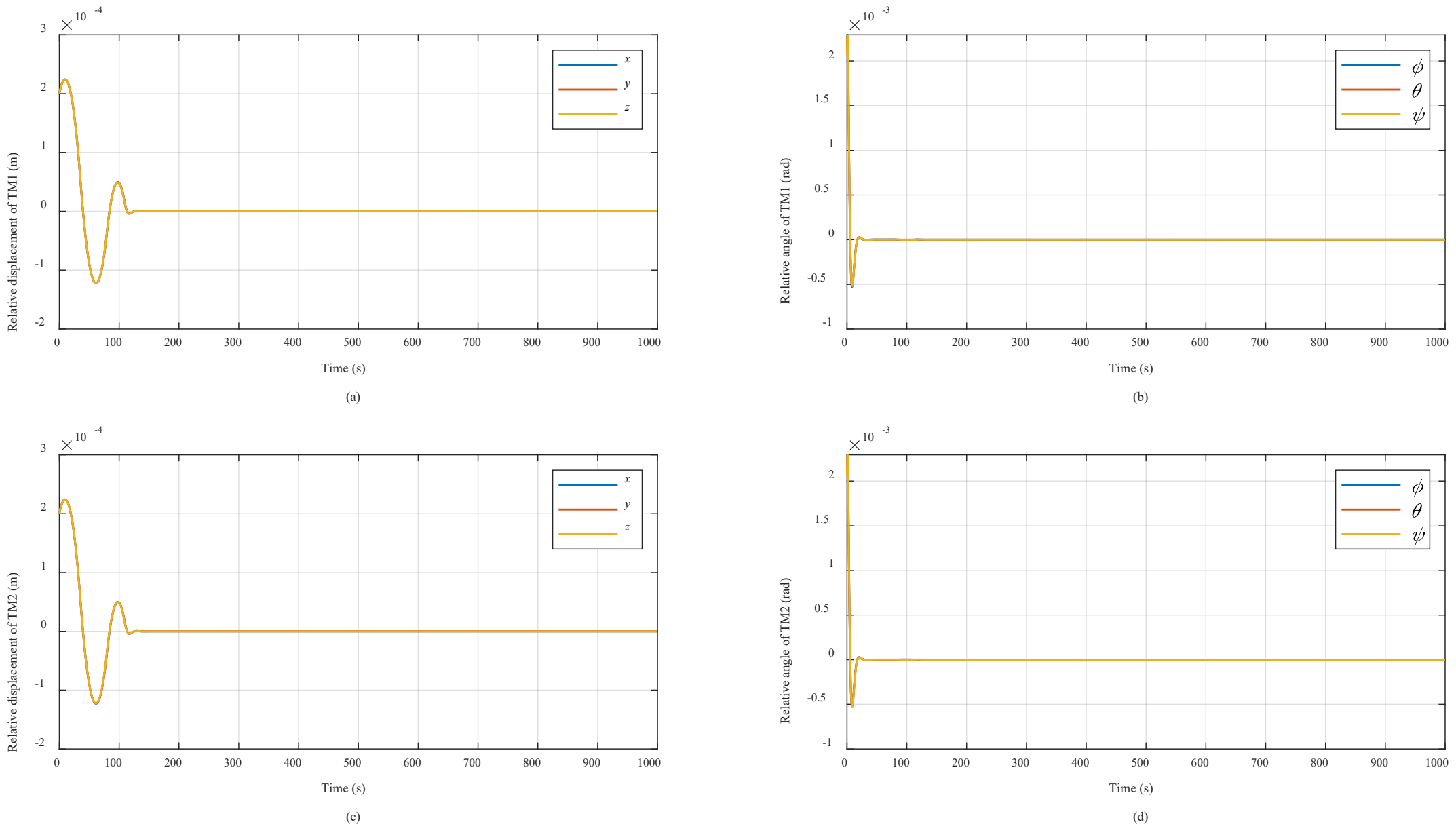

**Figure 4.** The relative displacements and angles of the TMs

As shown in the Figure 3 and 4, the simulation results indicate that the attitude attitude and the MOSA rotation angle eventually converge to their desired values. The relative pose of the TM rapidly converges to zero without any collision with the cage. These results demonstrate that the SINDy-based MPC successfully accomplishes the TM capture task.

## 5. Conclusions

In this paper, the nonlinear dynamics of drag-free satellite are identified based on Koopman operator theory and SINDy algorithm. The polynomial function of the original state is taken as the nonlinear observation, and the finite-dimensional approximation of Koopman operator is obtained by the SINDy. MPC based on Koopman linearization model avoids solving complex nonconvex problems, and greatly improves the calculation efficiency. A linear MPC controller is designed to control the drag-free satellite. The simulation results show that the SINDy-MPC method can effectively control the drag-free satellite. The results of this paper show that the general modelling and control methods based on Koopman operator theory, such as SINDy, have a good prospect in modelling and controlling complex satellite, but further research is still needed to improve the performance of these methods.